# Dislocation interaction with a tilt low angle grain boundary in bi-crystal SrTiO$_3$

Kuan Ding[1a], Atsutomo Nakamura[2], Patrick Cordier[3,4*], Xufei Fang[2,5b*]

[1]Department of Materials and Earth Sciences, Technical University of Darmstadt, 64287 Darmstadt, Germany

[2]Department of Mechanical Science and Bioengineering, Osaka University, 1-3 Machikaneyama-chou, Toyonaka, Osaka, 560-8531, Japan

[3]Unité Matériaux et Transformations, Université de Lille, 59655 Villeneuve d'Ascq Cedex, France

[4]Institut universitaire de France (IUF), 75005 Paris, France

[5]Institute for Applied Materials, Karlsruhe Institute of Technology, Kaiserstr. 12, 76131 Karlsruhe, Germany

[a]*Currently at Max Planck Institute for Sustainable Materials, Max-Planck-Straße 1, 40237 Düsseldorf, Germany*

[b]*Previously at affiliation [1] Department of Materials and Earth Sciences, Technical University of Darmstadt, 64287 Darmstadt, Germany*

*Corresponding authors: patrick.cordier@univ-lille.fr (P.C.); xufei.fang@kit.edu (X.F.)

**Abstract**

For potentially wider applications of ceramics with dislocation-tuned mechanical and functional properties, it is pertinent to achieve dislocation engineering in polycrystalline ceramics. However, grain boundaries (GBs) in general are effective barriers for dislocation glide and often result in crack formation when plastic deformation in ceramics is attempted at room temperature. To develop strategies for crack suppression, it is critical to understand the fundamental processes for dislocation-GB interaction. For this purpose, we adopt here a model system of bi-crystal SrTiO$_3$ with a 4° tilt GB, which consists of an array of edge dislocations. Room-temperature Brinell indentation was used to generate plastic zone at mesoscale without crack formation, allowing for direct assessment of GB-dislocation interaction in bulk samples. Together with dislocation etch pits imaging and transmission electron microscopy analysis, we observe dislocation pileup, storage, and transmission across the low angle tilt GB. Our experimental observations reveal new insight for dislocation-GB interaction at room temperature at mesoscale.

**Keywords:** dislocation; low angle tilt grain boundary; slip transmission; SrTiO$_3$; TEM

# 1. Introduction

Dislocations in ceramics have been proved promising for tuning functional and mechanical properties [1-10]. In order to engineer dislocations into ceramic materials, mechanical deformation has been an effective way that attracts increasing attention, particularly at room temperature [11]. Dislocation-mediated plasticity has been proven feasible in bulk single-crystal oxides at either high temperature [4, 7, 12] or room temperature [13-15], making single crystals an appealing candidate to investigate dislocation-tuned functionality. However, polycrystalline ceramics are commonly used in applications, with much more cost-effective processing and fabrication compared to single crystals. Therefore, mechanical tailoring of dislocations in polycrystalline holds great potential for extending dislocation-tuned functionality into real applications.

Grain boundaries (GBs) in polycrystalline materials generally act as effective barrier for dislocation motion. Particularly for ceramics, the insufficient independent slip systems at room temperature [16] do not fulfill the von Mises or Taylor criterion, which suggests five independent slip systems are required for arbitrary plastic deformation [17], making it challenging to engineer dislocations without crack formation using the mechanical deformation approach at the macroscale. For instance, tensile testing of polycrystalline MgO at room temperature showed that the dislocations were initiated at the vicinity of the GB, but no slip transmission across the GB was observed [18]. Four-point bending tests of polycrystalline LiF showed that GBs were strong barriers to dislocation slip [19]. Bulk compression test of NaCl polycrystals at room temperature unveiled that low-angle GB (LAGB) allowed transmission of gliding dislocations, whereas high-angle GB (HAGB) acted as an impenetrable obstacle to gliding dislocations [20]. These earlier studies provided valuable insights into dislocation-GB interaction as well as crack formation during bulk deformation, yet the dislocation microstructures were not characterized due to the lack of advanced characterization techniques back then.

In order to gain a more detailed understanding of the interaction between dislocations and specific grain boundaries, recent studies have been carried out at nanoscale in bi-crystal oxide ceramics. Kondo et al. [21] performed *in situ* nanoindentation in a TEM (transmission electron microscope) and directly observed the interaction of individual dislocation with a 1.2° tilt LAGB as well as a Σ5 HAGB in bi-crystal $SrTiO_3$. The processes with dislocations first impeded and then transmitted through this LAGB were visualized for the first time. In stark contrast, dislocation pileup and no transmission were found at the Σ5 HAGB. In a most recent study, the same group of Ikuhara's observed dislocation transmission in twist LAGB as well as jog formation caused by the interaction between the grain boundary screw dislocations (as fabricated) and the incoming dislocations (induced by mechanical loading) [22]. These *in situ* studies have advanced our understanding of the dislocation-GB interaction at the microscopic scale. However, in order to observe the dislocation structures in TEM, thin foil with hundreds of nanometers must be used, resulting in different deformation boundary conditions and stress states than in the bulk. The question remains as to what extent the deformation mechanisms obtained from *in situ* TEM deformation could be directly transferred or correlated with bulk deformation mechanisms [23]. In another study, Nakamura et al. [24] adopted bulk bi-crystals and investigated the nanoindentation response of $ZrO_2$ with different types of high-angle symmetrical tilt boundaries and $SrTiO_3$ with a symmetrical Σ5 boundary. They showed dislocation pileup, penetration, or generation through GBs by TEM characterization. Although the tests were carried out in bulk samples, the nanoindentation tests were limited at nano-/microscale. The size effects as well as the high degree of local confinement in nanoindentation tests limits this approach to represent the plastic deformation behavior at meso/macroscales.

To fill the gap in the length scale, Okafor et al. [25] recently adopted cyclic Brinell indentation with a millimeter-sized spherical indenter, and achieved up to a dislocation density of ~$10^{13}$ m$^{-2}$ in crack-free plastic zones with hundreds of micrometers in single-crystal $SrTiO_3$ at room temperature. This approach is simple and straightforward in the sense that it allows the study of dislocation-GB interaction at the mesoscale scale up to bulk scale, relevant for functional and mechanical testing. Later, Okafor et al. [26] applied this method to polycrystalline coarse-grained $SrTiO_3$ and achieved near-surface plastic deformation without crack formation. They made use of the samples free surface to relax the von Mises or Taylor criterion. However, a detailed dislocation-GB interaction at mesoscale was not attempted [26]. Here, we adopt the Brinell indentation method [25] to engineer dislocations and make them interact with GB, using a bi-crystal $SrTiO_3$ samples containing a 4° tilt LAGB. The dislocation-GB interaction was first revealed at the surface by dislocation etch pits analysis. Furthermore, we perform a transmission electron microscope (TEM) investigation of these specimens to gain a more in-depth understanding of the dislocation-GB interaction underneath the surface.

## 2. Materials and Methods

### 2.1. Bi-crystal fabrication

The $SrTiO_3$ (STO) bi-crystal with a [001]/(110) 4° tilt grain boundary (**Fig. 1A, B**) was fabricated using the thermal diffusion bonding technique [27]. Two STO single crystals with 2° off (110) surfaces were polished to mirror-like surfaces. Afterwards, the surfaces were cleaned with ethanol, and the two single crystals were attached to each other to create a 4° tilt grain boundary, with the two crystals having a +2° and -2° inclination from the (110) surface around the [001] direction. The bicrystal used in this study was prepared by diffusion bonding at 1000 °C for 10 h under a uniaxial load of 10 N. Further annealing at 1500 °C in air for 10 h was performed to achieve sufficient atomic diffusion [28].

### 2.2. Sample surface preparation

The sample surface (perpendicular to the GB plane, see **Fig. 1C**) was metallographically prepared to obtain a smooth surface without inducing additional surface dislocations. The surface was mechanically ground using SiC based grinding papers with grade from P1200 to P4000. The sample was then mechanically polished with diamond particles with sizes of 6, 3, 1, and 0.25 µm, respectively. The final polishing was carried out with colloidal silica polishing suspension OP-S (Struers, Germany) for ca. 15 hours. Afterwards, the sample surface was cleaned with distilled water and then dried with ethanol.

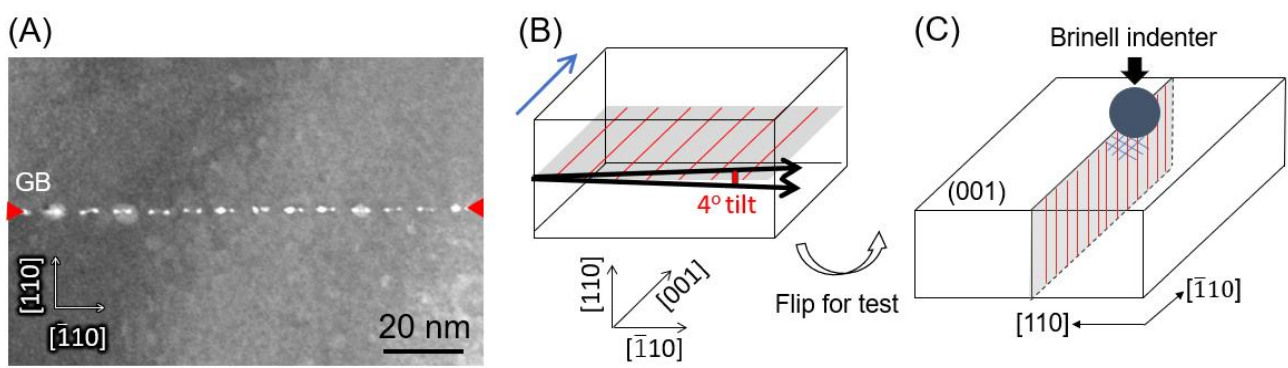

*Fig. 1 (A) Bright-field TEM image of the as-fabricated 4° tilt low angle grain boundary (GB, as indicated by the red triangles); (B) Schematic of the bi-crystal fabrication and crystal orientation, with the GB plane (110) highlighted in grey, and the edge dislocations as red lines; (C) Schematic illustration of the Brinell indentation tests in one grain to induce dislocations without crack formation, allowing for dislocation-GB interaction studies. The patched lines underneath the indenter indicate the generated slip traces.*

## 2.3. Brinell indentation

Brinell indentation tests [25] were performed using a universal hardness testing machine (Karl-Frank GmbH, Germany) mounted with a spherical indenter (hardened steel) with a diameter of 2.5 mm. A test load of 1.5 kg was used to generate dislocations without crack formation, allowing for direct investigation of the dislocation-GB interaction (**Fig. 1C**).

## 2.4. Microstructural characterization

Optical images of the sample surface after indentation were captured using a Zeiss Axio Imager 2 optical microscope (Zeiss, Germany) with circular differential interference contrast (C-DIC) and dark field mode. Dislocation etch pits study was carried out using chemical etching method by immersing the sample in 50% $HNO_3$ containing 16 drops of HF solution for 50-60 seconds. The etched sample surface was characterized using a LEXT OLS4100 laser confocal microscope (Olympus, Japan). A thin layer of carbon was sputtered onto the sample surface before SEM characterization to reduce the surface charging. The SEM micrographs were taken in a Tescan MIRA 3 XMH SEM (Tescan, Czech Republic) with an accelerating voltage of 5 kV.

## 2.5. TEM characterization

Before deformation, the as fabricated grain boundary was examined (**Fig. 1A**) using TEM. Before deformation, thin foils for TEM observation were prepared from the above-fabricated 4°-tilt low angle grain boundary. The samples were initially sliced to include the grain boundary plane, followed by Ar$^+$ ion milling to achieve electron transparency. Dislocation arrays along the grain boundary were then examined using an ultra-high-voltage transmission electron microscope (JEOL JEM-1000k RS) operated at 1000 kV. After deformation, TEM characterizations were carried out using the electron microscopy facility of the Advanced Characterization Platform of the Chevreul Institute (University of Lille), with a FEI® Tecnai G220Twin microscope, operating at 200 kV equipped with a LaB$_6$ filament. The TEM thin sections were prepared by focused ion beam (FIB) with a FIB/SEM ZEISS Crossbeam 550.

## 3. Results and Analyses

As illustrated in the optical images in **Fig. 2**, large plastic zones (~200 μm in diameter for the indent imprint) without crack formation have been achieved using Brinell indentation. This mesoscale approach is verified to be feasible, as multiple indents were performed with different distances to the GB for optimizing the testing conditions, giving excellent reproducibility in the plastic zones without crack formation. Here in **Fig. 2B**, two indent imprints are highlighted, with the corresponding dark-field images (**Fig. 2C**) corroborating no visible crack formation along the GB. Slip bands are observed that terminate on (#2 in **Fig. 2B**) or transmit (#3 in **Fig. 2B**) across the LAGB.

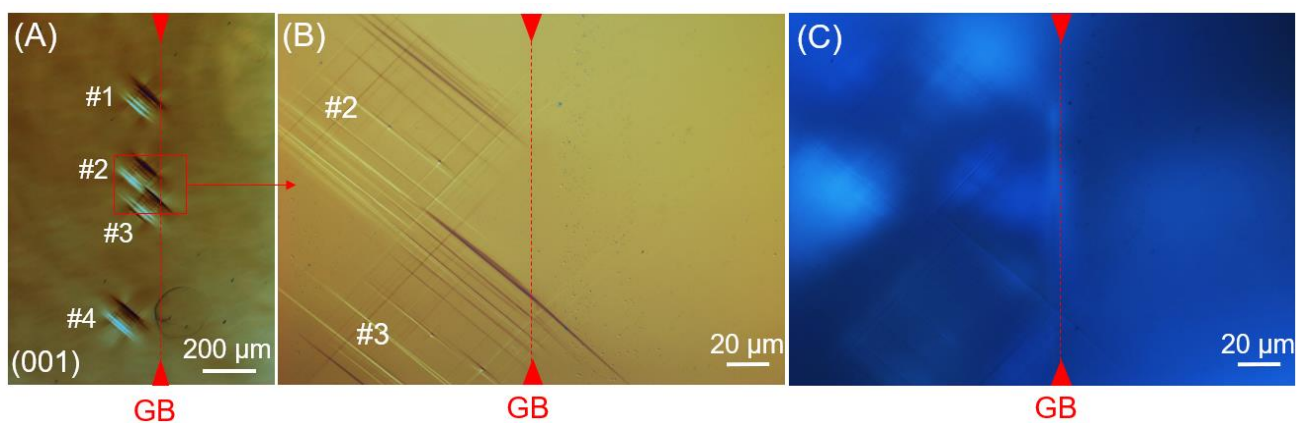

*Fig. 2. Optical images of representative Brinell indents near the grain boundary (GB). (A) Multiple indents were performed with different distances to the GB for optimizing the experimental parameters; (B) Zoom-in region showcasing the slip traces terminate (#2) or penetrating (#3) the GB; (C) Corresponding dark-filed image of the region in (B) revealing no crack formation at the GB.*

After optical image examination, the (001) surfaces were chemically etched to reveal the dislocation etch pits (**Fig. 3**). The etched GB is identified as the vertical straight line across the sample surface (**Fig. 3A,** yellow triangles). The Brinell indent was imprinted next to the GB on the right-side grain, with a load of 1.5 kg and 30 cycles. The higher number of cycles is used for this indent to generate sufficient dislocations to reach and penetrate the GB. The dense etch pits on the right side of the GB in **Fig. 3A** correspond to dislocations generated within the Brinell imprint.

As revealed by the surface etch pits (**Fig. 3A**), this 4° tilt LAGB can already effectively impede most of the dislocations. Several slip bands in the [$0\bar{1}0$] and [$\bar{1}00$] directions successfully transmitted (indicated by the white arrows in **Fig. 3A**) across the grain boundary into the adjacent grain. Two of these transmission sites were highlighted in **Fig. 3B1** and **C1**. To obtain the in-depth information on the dislocations-GB interactions at these two selected sites, TEM lamella lift-outs (**Fig. B2** and **C2**) were performed parallel to the (010) planes. Note the yellow lines in **Fig. B1** and **C1** indicate the intersection of the (010) planes with the sample surface (001)). The overview of the two TEM lamellae can be found in the **Supplementary Materials** (**Fig. S1-S2**).

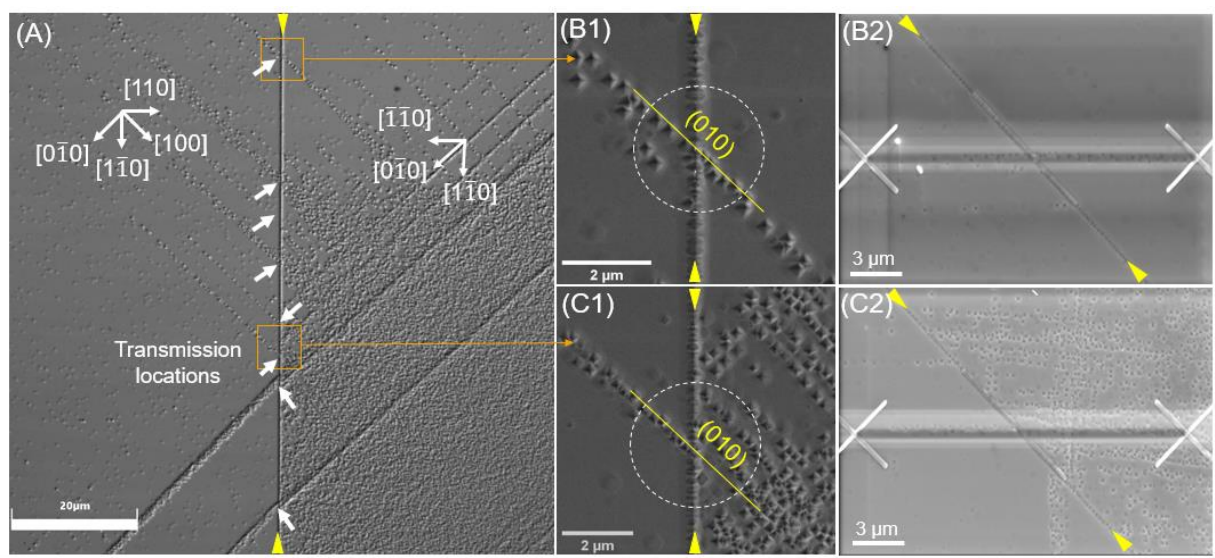

*Fig. 3 (A) Laser microscope image of dislocation-GB interaction from a Brinell indentation imprint (on the right grain) after chemical etching. The slip transmission sites are indicated by the white arrows. (B1-B2) SEM micrographs for the zoomed-in region for one slip transmission event. The X-X indicate the corresponding TEM lamella lift-out position for the same location, as B2 is 45° rotated with respect to B1; (C1-C2) SEM micrographs for the zoomed-in region for another slip transmission event. The X-X indicate the corresponding TEM lamella lift-out for in-depth observation, with C2 45° rotated with respect to C1. The GB is indicated by the yellow triangles in all cases.*

For the first selected slip transmission site (**Fig. 3B**), the GB region has been successfully captured in the TEM analysis, as indicated by the yellow triangles in **Fig. 4**. The black stripe on the right to the GB is the thick sample region retained during FIB milling (see also the overview in **Fig. S1**) for supporting the thin TEM lamella to minimize bending. In these dark-field TEM images, the dislocation lines are visualized by the bright line contrasts beneath the sample surface. The surface etch pits (green arrows) were successfully captured in **Fig. 4**, with each etch pit tailed by a dislocation. These dislocations are in contrast with **g**: 002 (**Fig. 4**) and 011 (**Fig. S3**) and out of contrast with **g**: $01\bar{1}$ (**Fig. S4**). They exhibit short segments compatible with inclusion in a plane inclined at 45° to the thin section. We therefore identify these dislocations as [011] dislocations gliding in $(01\bar{1})$. **Figure S3** even shows that these dislocations are mostly oriented screw, in agreement with the previous dislocation etch pit analysis by Javaid et al. [29] using nanoindentation tests. On the left grain, etch pits tailed by dislocations (yellow arrows at the surface) were also captured, corresponding to the transmitted dislocations as in **Fig. 3B1**, which are shown here to belong to the $[011](01\bar{1})$ slip system. This observation directly supports the validity of etch pit method for dislocation observation in SrTiO$_3$. Beneath the surface there are more dislocation lines observed (dislocation density higher than ~10$^{12}$ m$^{-2}$), suggesting that the total dislocation density in the deformed region should be higher than by merely counting the surface dislocation etch pits. Trapped dislocations are also present in the interface. The most frequent ones (marked in white) have orientations compatible with $[1\bar{1}\bar{1}]$. This was verified by observing the specimen under different orientations, close to the $[0\bar{1}1]$ (**Fig. S3**) and $[0\bar{1}\bar{1}]$ (**Fig. S4**) zone axes. Even though in one case (**Fig. S4**) the dislocations are out of contrast, the orientations of the lines are compatible with the $[1\bar{1}\bar{1}]$ direction, which is at the intersection between the LAGB (110) plane and that of the [011] dislocations: $(01\bar{1})$. Therefore, we interpret these lines as [011] dislocations trapped in the grain boundary interface. It should be noted that the LAGB shows other contrasts of dislocations. One of these is tentatively indicated by $[1\bar{1}0](110)$ in **Fig. 4**, although it has not been fully characterized. These marginal observations will not be discussed further.

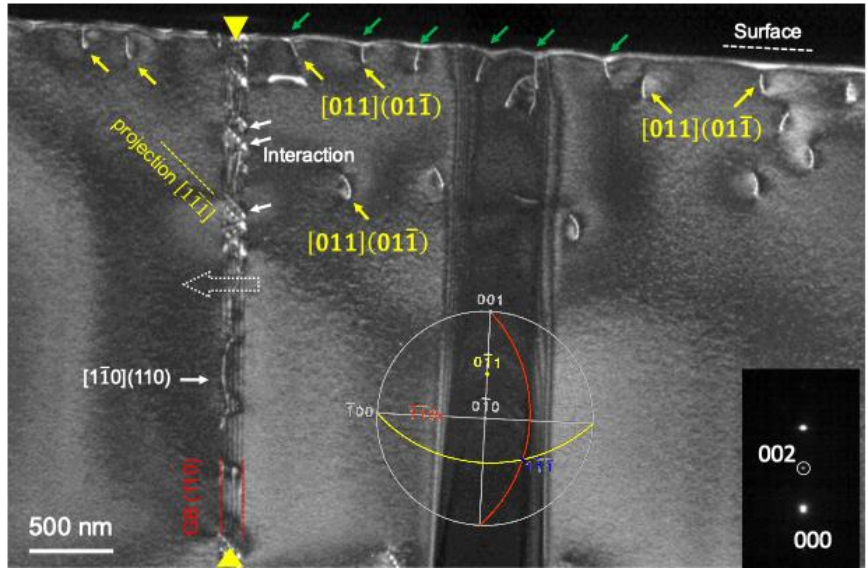

*Fig. 4. Weak-beam dark-field (g/2g with g: 002) TEM observation of location 1 in **Fig. 3B**. Overview of the dislocation-GB interaction with dislocations intersecting the free surface as well as the GB (thick yellow triangles).*

**Figures 5 and 6** present a second TEM thin section which was FIB-milled from the interaction site in **Fig. 3C.** The white dashed arrow in **Fig. 5** indicates the direction of the dislocations travel. Here we observe more dislocations generated on the right grain next to the GB (**Fig. 5**). Transmitted dislocations across the GB is also observed at the surface (yellow arrows in **Fig. 5**). Some dislocations in the right grain have all the characteristics of those analyzed in **Figs. 4, S3-4**. We therefore identify them as $[011](01\bar{1})$ dislocations (marked in yellow). This identification is further supported here by the fact that they are out of contrast with **g**: 200 (**Fig. S5**). Another family of dislocations (marked in blue) is in contrast with **g**: 011 (**Fig. S6**), 200 (**Figs. 5, 6 and S5**), and 002 (**Figs. 6 and 7**). They are also contained in the $(10\bar{1})$ plane, which is edge-on in **Fig. 6**. Therefore, we interpret these dislocations as belonging to the $[101](10\bar{1})$ slip system. These dislocations give rise to another family of lines trapped in the interface with a different orientation from that of the $[011](01\bar{1})$ dislocations. Observation of the specimen near the zone axes $[010]$ (**Figs. 5 and 6**), $[0\bar{1}\bar{1}]$ (**Fig. S5**) and $[0\bar{1}1]$ (**Fig. S6**) allows the direction of these lines to be identified as $[\bar{1}1\bar{1}]$, which is indeed the intersection between the (110) and $(10\bar{1})$ planes. Furthermore, **Figure 7** demonstrates that both $[011](01\bar{1})$ and $[101](10\bar{1})$ dislocations are eventually transmitted across the LAGB into the left grain.

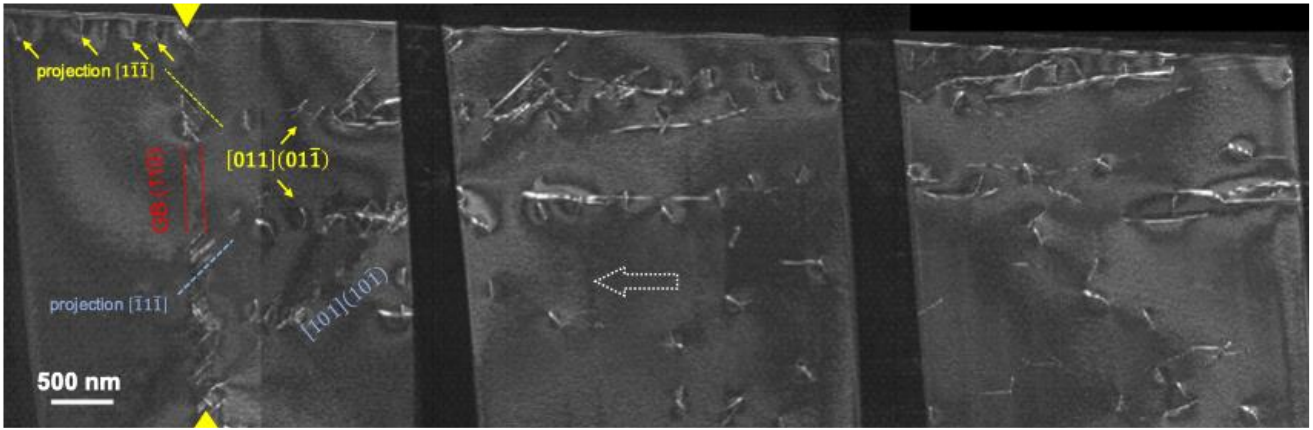

*Fig. 5.* TEM lamella No. 2. Weak-beam dark-field (g/2g with **g**: 200) TEM mosaic micrographs showing the dislocation microstructure below the surface. The LAGB location is highlighted by the yellow arrows on the left. The white dashed arrow indicates the travel direction of the dislocations induced by the Brinell indentation. More dislocations are observed to interact with the GB leading to two distinct directions of lines of entrapped dislocations.

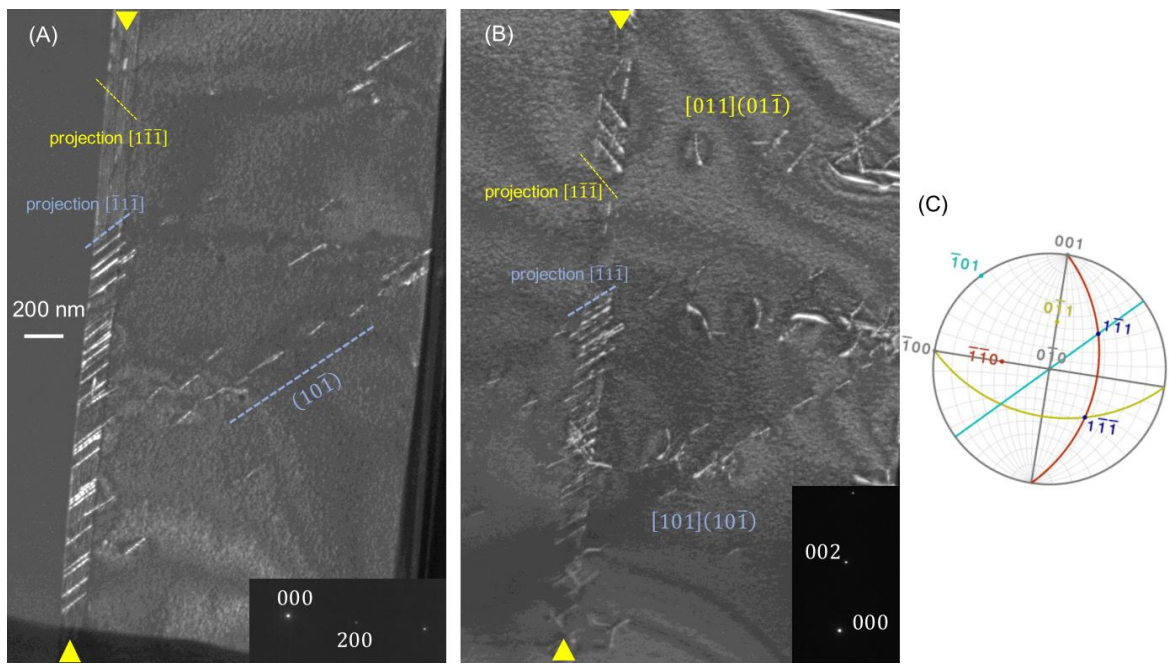

*Fig. 6.* TEM lamella No. 2, with region near the LAGB showing dislocations heading towards it. (A) Weak-beam dark-field (g/2g with **g**: 200) close to the $[0\bar{1}0]$ zone axis (see stereoplot in subfigure C). (B) Weak-beam dark-field (g/2g with **g**: 002) close to the $[0\bar{1}0]$ zone axis. The line directions of dislocations entrapped in the LAGB are parallel to $[1\bar{1}\bar{1}]$ (yellow) and $[\bar{1}1\bar{1}]$ (blue). $[011]$ dislocations (yellow) are out of contrast in (A). The glide plane of $[101]$ dislocations (blue) is seen edge-on.

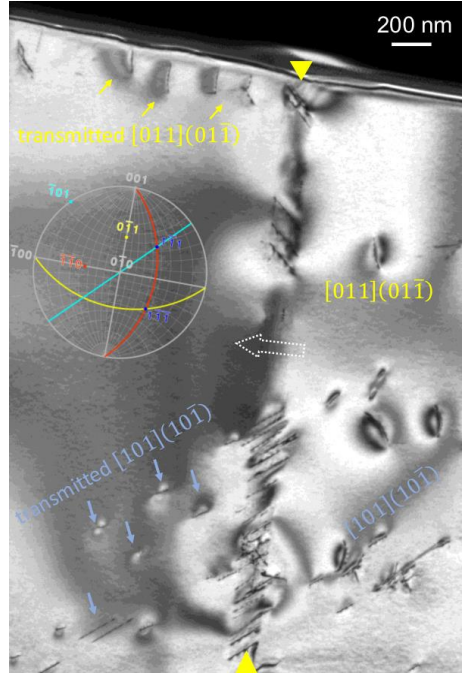

***Fig. 7.*** *TEM lamella No. 2, with the same region as in Fig. 6. Dark-field with **g**: 002 close to the $[0\bar{1}0]$ zone axis. The diffraction conditions are adjusted to highlight dislocations on both sides of the LAGB. Both $[011]$ and $[101]$ dislocations are transmitted.*

## 4. Discussions

### 4.1. Dislocation-GB interaction

The results in **Figs. 3-7** reveal complex dislocation-GB interaction processes even for a simple tilt LAGB. Besides dislocation pile up and transmission, it appears that dislocation storage also occurs at the GB for both $[011](01\bar{1})$ and $[101](10\bar{1})$ slip systems activated. The common criteria to describe the feasibility for slip transmission is the *m*' factor [30]: $\boldsymbol{m}' = \cos\phi\cos\kappa,$ where $\phi$ is the angle between the slip plane normal, and $\kappa$ is the angle between the slip directions. The absolute value of *m*' is between 0 and 1, with *m*' being close to 1 for easy slip transmission. The calculated *m'* value for LAGB is 0.9976, suggesting slip transmission is easier to occur for the LAGB. The possible dislocation-GB interaction is illustrated **Fig. 8**. As illustrated in the experimental observation (**Figs. 4-5**), it is evident that there is partly transmission of the dislocations but also partly dislocation reaction and/or dislocation storage in the LAGB. In what follows we discuss these possible scenarios.

As illustrated in **Fig. 1**, the LAGB has a plane of (110), and is made up of an array of edge dislocations with $[1\bar{1}0]$ Burgers vector. It is likely that storage is a consequence of reactions between the GB dislocations and the incoming screw dislocations. In case of $[011](01\bar{1})$ incoming dislocations, the reaction $[011] + [1\bar{1}0] = [101]$, is energetically favorable, inducing a tendency to storage. As for the dislocations $[101](10\bar{1})$, the reaction would be $[101] + [1\bar{1}0] = [2\bar{1}1]$ which is not favorable according to Frank's energy criterion [29]. However, this analysis considers perfect dislocations instead of dissociated dislocations, with the latter being the case for STO at room temperature [31, 32]. Consider the stacking fault energy of STO is high (~136 mJ/m$^2$ for room temperature [31]) and the partials are spread a few unit cells apart [32, 33], leading to more complex situations with extended nodes as illustrated in Besson et al. (1996). We expect future work with more detailed but strenuous characterization with advanced TEM shall provide more detailed information to confirm dislocation reactions, preferably coupled with atomistic simulations, which is beyond the current scope of the work.

It is also worth noting that Kondo et al. [21] directly visualized transmission of screw dislocations across a LAGB (with the GB dislocations having a Burgers vector of [100], different from the current case) in bi-crystal STO in their *in situ* TEM indentation tests. The differences are also reflected in the transmitted dislocations, which exhibited newly formed superjog segment with a Burgers vector of [1$\bar{1}$1] according to the reaction of $\mathbf{b}_{Jog} = \mathbf{b}_{Lattice} + \mathbf{b}_{GB} = [0\bar{1}1] + [100] = [1\bar{1}1]$. Nevertheless, LAGBs in STO seem to allow for easy slip transmission regardless of their GB configuration.

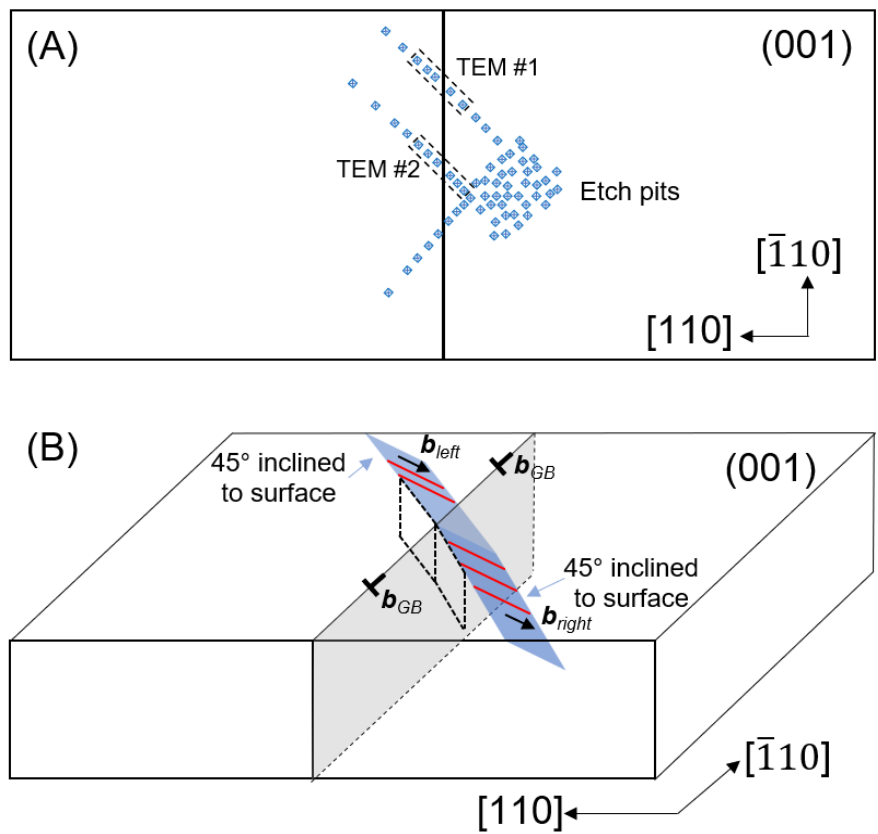

*Fig. 8* (A) 2D plan view of the indented area near GB, with the two TEM lamellae lift-outs indicted. The blue squares in the right grain indicate the dislocation etch pits, with the activated slip planes {110} being 45° inclined to the sample (001) surface. The dislocation lines, indicated in red, are predominantly screw types. (B) 3D perspective of a representative slip plane with dislocation transmission into the adjacent

*grain. These dislocation lines, when projected to the TEM lamellae planes (black dashed rhombus), will be 45° inclined.*

As schematically indicated in **Fig. 8**, the dislocations induced by Brinell indentation are predominantly screw types. This has been reported and confirmed elsewhere [21, 25, 34, 35]. The slip planes activated are of {110} types, when projected on the TEM samples plane, 45° inclined dislocation line contrasts will be formed, as observed in **Figs. 4-5**. In particular, there are 4 equivalent {110} planes that can be activated, with a 45° inclined angle to the sample surface. This means that for TEM lamella 1 (in **Fig. 4**), it is likely that only one of these 45° slip plane was activated, corresponding to the singe slip trace and single array of dislocation etch pits in **Fig. 3B1**. For TEM lamella 2 (**Fig. 5**), the multiple, mutually perpendicular slip traces as well as the arrays of dislocations etch pits suggest more than one of such 45° slip planes have been activated. The projections along the TEM lamella hence will generate not only 45° but also 135° (or minus 45°) inclined dislocation lines with respect to the sample surface, as confirmed by the TEM observation in **Fig. 5A**.

The discussions based on **Fig. 8** raises question on the applicability of conventional models using pure geometric mismatch-based slip transfer metrics in predicting the slip transfer in oxides as these models are only formulated in terms of the misorientation angles by simply assuming a transmission can happen when $\theta < 15º$. The local atomistic structure at the slip-GB intersection as well as the activated slip systems that are interacting with the GB plays an important role in determining whether a transmission will occur or not. If the local structure, such as the one corresponding to a GB dislocation, carries the same Burgers vector as that of the incoming dislocation, it will introduce a strong repulsive force acting on the incoming dislocation. A pileup can still form due to such a strong repulsive force, even the misorientation angle across the GB is as low as $\theta = 4º$. Furthermore, the relatively higher Peierls barrier [36] and high stacking fault energy [31, 37] leads to the low mobility of screw dislocations, resulting in limited number of transmitted dislocations. Moreover, Kondo et al. [21] proposed that the lattice screw dislocations can react with the grain boundary edge dislocations to form jogs when crossing the LAGB. Using dark-field TEM imaging, they observed shift in the GB edge dislocation lines with super-jogs formed on the GB dislocations as well as kinks on the lattice dislocations. Trapped or stored lattice dislocations were also observed on the GB plane.

### 4.2. Crack suppression using Brinell indentation

A critical aspect to account for during room-temperature plastic deformation in ceramics is the cracking. Considering the size-dependent competition between cracking and plasticity in ceramics deformation, it is pertinent to illustrate more details on the choice of Brinell indentation for investigating dislocation-GB interaction. Even for SrTiO$_3$ [13] and MgO [38], the two well-known oxides that exhibit excellent room-temperature dislocation plasticity in bulk deformation, cracks formation can still be easily triggered. In the indentation case, due to the confinement and local hydrostatic compressive stress, dislocation plasticity is favored as in the case of Brinell indentation [15, 25, 26] (see also **Fig. 2**). Nevertheless, a strong size-dependent competition between the dislocation plasticity and crack formation has been recently identified [39], which concerns primarily the incipient plasticity as well as the crack formation due to the dislocation plie-up and dislocation interaction [40] underneath the indenter. To be specific, smaller indenter may favor dislocation nucleation, yet cracks are also more easily to form due to the dislocation pileup and interaction that occurs more easily due to the much higher stress level and higher degree of dislocation multiplication. This is evidenced by e.g., sharp Berkovich indenter [41] or spherical indentation with smaller indenter tips [35]. The formation of cracks as well as the confined dislocation distribution induced by smaller indentation result in complications for dislocation-GB interaction, as illustrated in **Fig. 9,** where cracks may proceed prior to dislocation-GB interaction. In this regard, lowering the stress level by using larger indenter is beneficial in crack suppression. This works for the oxides such as SrTiO$_3$, MgO, LiF, and many more ceramics [42] that exhibiting good dislocation mobility at room temperature. This also merits the general applicability of the Brinell indentation and scratching approach for exploring dislocation-GB interactions at mesoscale.

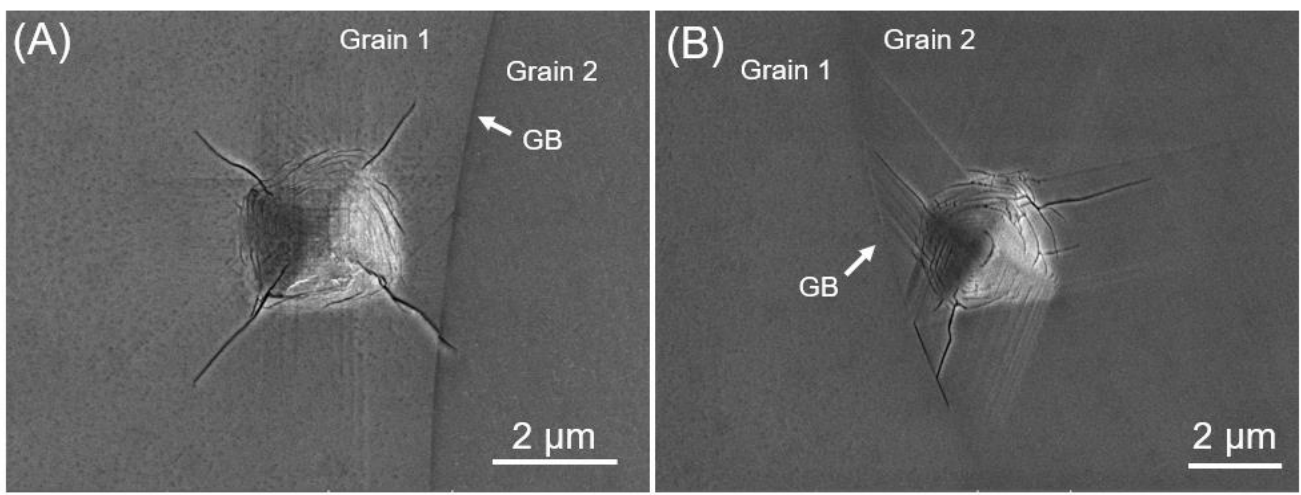

*Fig. 9 SEM micrographs of the indentation imprints left by using nanoindentation with an effective indenter tip radius of 2 µm at maximum load of 75 mN. Both indent imprints reveal slip traces accompanied by crack formation, making the pure dislocation-GB interaction analysis challenging. Note that the sample was chemically etched to reveal the dislocation etch pits, and the tests were performed on a coarse-grained SrTiO$_3$ to allow for probing different GBs.*

## 5. Conclusions

Room-temperature Brinell indentation is adopted as a feasible approach to induce plastic zones up to hundreds of micrometers without crack formation in model perovskite oxide $SrTiO_3$, allowing for mesoscale assessment of dislocation-GB interactions in its bicrystal. The combined near-surface analysis by dislocation etch pits study as well as the in-depth information obtained by TEM analysis in the grain interior reveals that 4º tilt LAGB, although being rather simple as an array of edge dislocations, exhibits yet very complex dislocation-GB interaction including trapping, storage, and slip transmission. This mesoscale indentation approach, when combined with scratching test, allows for probing multiple and different types of grain boundaries, and can help to pave the road for high-throughput analysis of dislocation-GB interactions at meso-/macroscale.

**Acknowledgements:**

X. Fang acknowledge the financial support by the European Union (ERC-Starting Grant, Project MECERDIS, grant No. 101076167). P. Cordier is supported by the European Research Council (ERC) under the European Union's Horizon 2020 research and innovation program under grant agreement No 787198 – TimeMan. FIB preparations are supported by the French RENATECH network, the CPER Hauts de France project IMITECH and the Métropole Européenne de Lille. Views and opinions expressed are however those of the authors only and do not necessarily reflect those of the European Union or European Research Council. Neither the European Union nor the granting authority can be held responsible for them. We thank Prof. K. Durst and Prof. J. Rödel at TU Darmstadt for access to the laser microscope, SEM, and optical microscope, as well as Prof. W. Rheinheimer at University of Stuttgart for providing the sample for us to perform the nanoindentation tests near the grain boundary.

**Author contribution:**

X.F. conceived the idea and designed the experiment. K.D. performed the indentation tests, collected the optical microscope and SEM images, and analyzed the data. A. N. fabricated the bicrystal and obtained the TEM data prior to mechanical deformation. P.C. collected the TEM data after deformation, analyzed the data and interpreted the results. X.F. and K.D. analyzed the data and wrote the first draft. All authors discussed the data and revised the draft.

**Conflict of interests:** The authors declare no conflict of interests.

**Supplementary Materials**

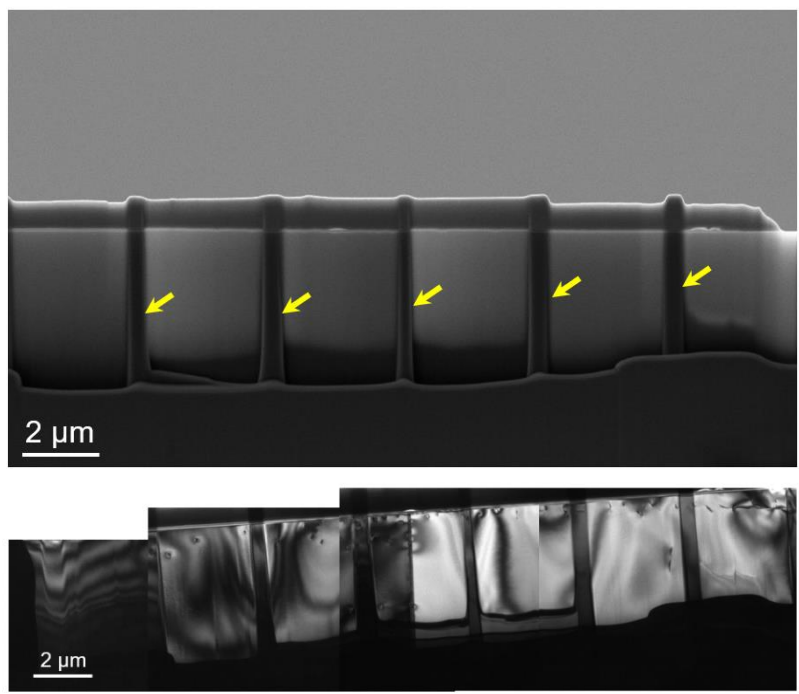

**Fig. S1.** TEM lamella No. 1 overview, correspond to region in **Fig. 3B.** Note that the ridges (yellow arrows) were purposely fabricated during FIB milling to support the thin, long TEM lamella to prevent bending.

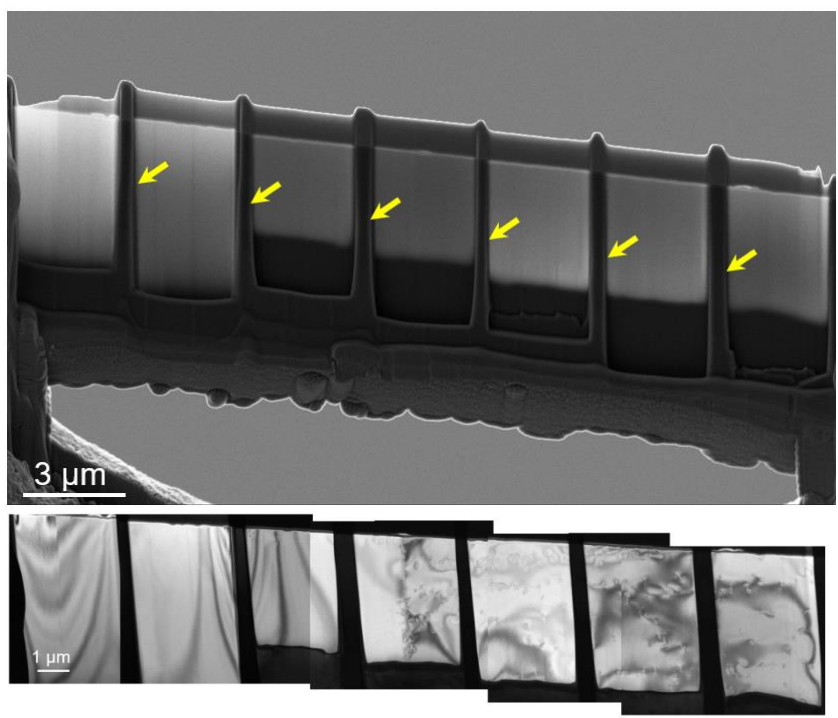

**Fig. S2.** TEM lamella No. 2 overview, correspond to region in **Fig. 3C.** Note that the ridges (yellow arrows) were purposely fabricated during FIB milling to support the thin, long TEM lamella to prevent bending.

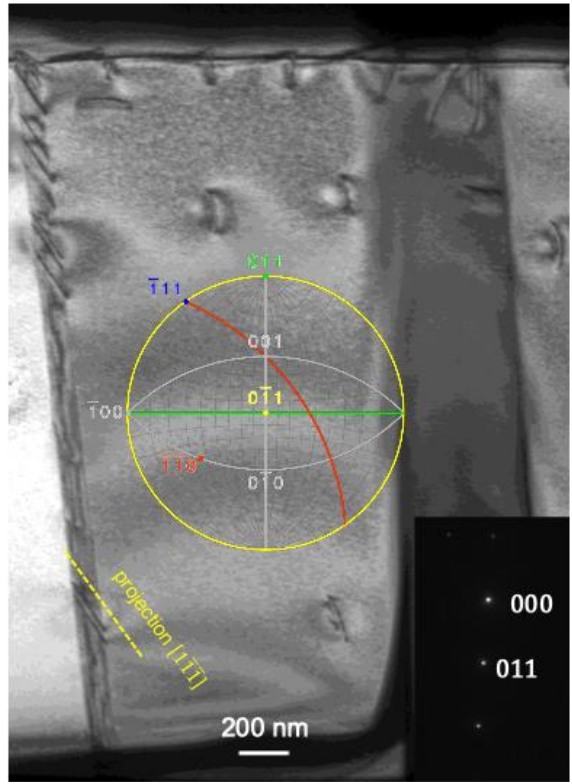

**Fig. S3.** TEM lamella No. 1 overview. Dark-field with **g**: 011 close to the $[0\bar{1}1]$ zone axis. The direction of dislocations entrapped in the LAGB is highlighted. They are parallel to the $[1\bar{1}\bar{1}]$ direction.

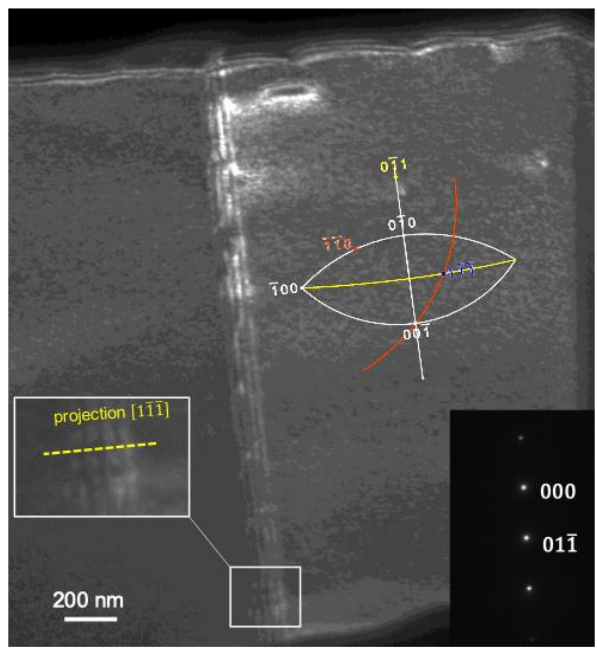

**Fig. S4.** TEM lamella No. 1 overview. Dark-field with **g**: $01\bar{1}$ close to the $[0\bar{1}\bar{1}]$ zone axis. The dislocations indicated in yellow in **Fig. 4** are out of contrast. The residual contrast of two dislocations entrapped in the LAGB is highlighted in the inset. Their direction is parallel to the projection of the $[1\bar{1}\bar{1}]$ direction.

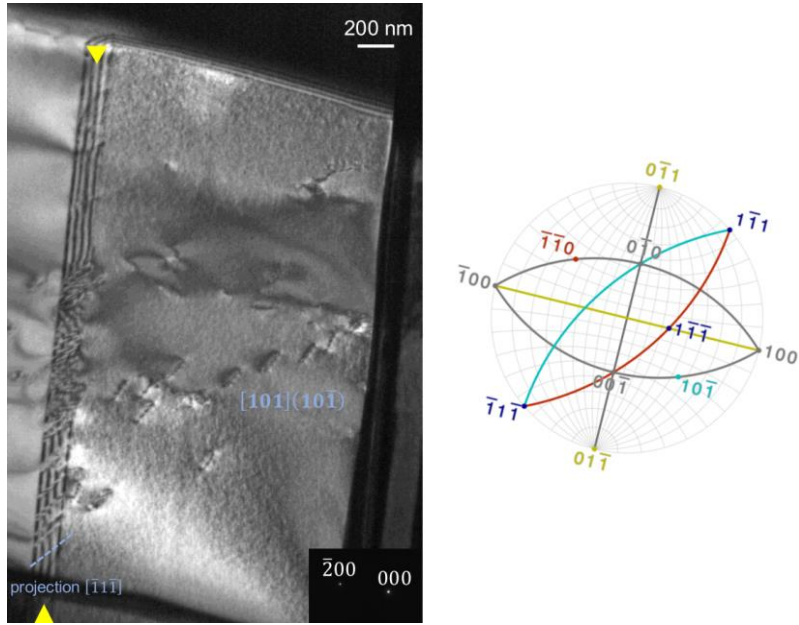

**Fig. S5** TEM lamella No. 2. Dark-field with **g**: $\bar{2}00$ close to the $[0\bar{1}\bar{1}]$ zone axis. [011] dislocations indicated in yellow in **Fig. 5** are out of contrast. The line directions of dislocations entrapped in the LAGB (and in contrast) are parallel to $[\bar{1}1\bar{1}]$.

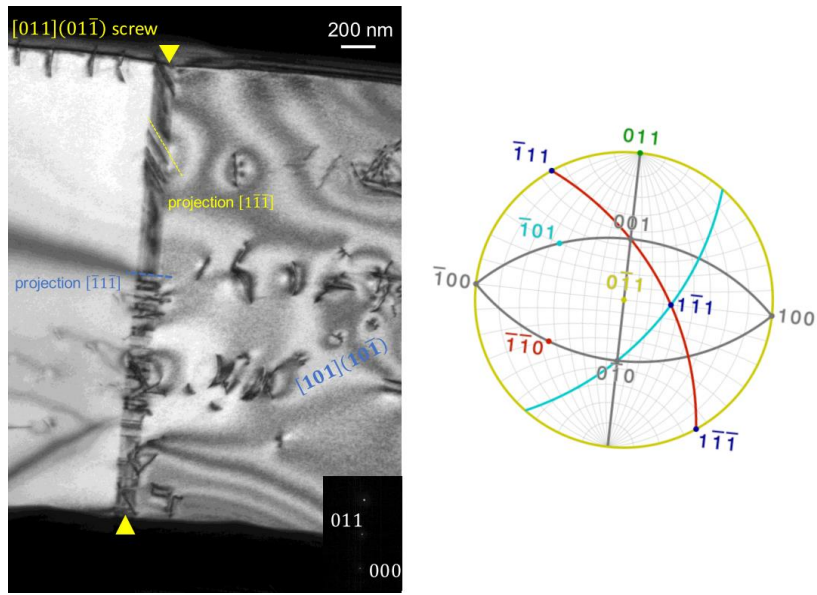

**Fig. S6** TEM lamella No. 2. Dark-field with **g**: 011 close to the $[0\bar{1}1]$ zone axis. The line directions of dislocations entrapped in the LAGB are parallel to $[1\bar{1}\bar{1}]$ (yellow) and $[\bar{1}1\bar{1}]$ (blue).